\documentclass[twocolumn,showpacs,preprintnumbers,amsmath,amssymb,floatfix]{revtex4}

\usepackage{graphicx}
\usepackage{dcolumn}
\usepackage{bm}
\usepackage{color}
\usepackage{url}

\newcommand{\be}{\begin{equation}}
\newcommand{\ee}{\end{equation}}
\newcommand{\ba}{\begin{eqnarray}}
\newcommand{\ea}{\end{eqnarray}}

\newcommand{\Ms}{M_{\odot}}

\def\ltsima{$\; \buildrel < \over \sim \;$}
\def\simlt{\lower.5ex\hbox{\ltsima}}
\def\gtsima{$\; \buildrel > \over \sim \;$}
\def\simgt{\lower.5ex\hbox{\gtsima}}
\begin{document}

\title{A Bayesian approach to the follow-up of candidate gravitational wave signals}

\author{J. Veitch$^1$, A. Vecchio$^{1,2}$} 
\affiliation{$^1$School of Physics and Astronomy, University of Birmingham, 
  Edgbaston, Birmingham B15 2TT, UK\\
  $^2$Northwestern University, Department of Physics and
  Astronomy, 2145 Sheridan Road, Evanston, IL 60208, USA
  }
 
\begin{abstract}
Ground-based gravitational wave laser interferometers (LIGO, GEO-600, Virgo and Tama-300) have now reached high sensitivity and duty cycle.
We present a Bayesian evidence-based approach to the search for gravitational waves, in particular aimed at the follow-up of candidate events generated by the analysis pipeline. We introduce and demonstrate an efficient method to compute the evidence and odds ratio between different models, and illustrate this approach using the specific case of the gravitational wave signal generated during the inspiral phase of binary systems, modelled at the leading quadrupole Newtonian order, in synthetic noise. We show that the method is effective in detecting signals at the detection threshold and it is robust against (some types of) instrumental artefacts. The computational efficiency of this method makes it scalable to the analysis of all the triggers generated by the analysis pipelines to search for coalescing binaries in surveys with ground-based interferometers, and to a whole variety of signal waveforms, characterised by a larger number of parameters.
\end{abstract}



\maketitle

\section{Introduction}
\label{s:intro}

Ground-based gravitational wave (GW) laser interferometers -- LIGO~\cite{BarishWeiss:1999}, Virgo~\cite{virgo}, GEO-600~\cite{geo} and TAMA-300~\cite{tama} -- have been in operation for a few years, alternating times of data taking for science analysis at progressively greater sensitivity and duty cycle, with commissioning periods to improve the instruments' performance. Recently LIGO has completed its fifth science run (S5) which lasted for $\approx 2$ years, recording one integrated year of data in triple coincidence between the two 4-km arm interferometers and the 2-km arm interferometer  at design sensitivity~\cite{ligoS5}. In addition, GEO-600 and Virgo were online for extended periods of S5. LIGO and Virgo are now being upgraded to the so-called enhanced configuration~\cite{eligo}, leading to a new science run (S6) in 2009 at a strain sensitivity a factor $\approx 2$ better than the present one. The much more intrusive upgrade to advanced LIGO/Virgo~\cite{advligo,advvirgo} is expected to lead to commissioning of the interferometers in 2014, and will increase the strain sensitivity by a factor $\approx 10$ across the whole frequency band and shift the low-frequency cut-off to about 10 Hz.
                                   
The search for GWs has therefore reached a stage in which the first direct detection is plausible and there are expectations that by the time the instruments operate in advanced configuration it will be possible to routinely observe a wide variety of sources in this new observational window and study them in detail, see~\cite{CutlerThorne:2000} and references therein.

One of the most promising classes of sources are binary systems of compact objects, which are reviewed in ~\cite{CutlerThorne:2000}. GW laser interferometers are omni-directional detectors that continuously monitor the whole sky looking for rare and/or weak signals. The general approach to the analysis of the data is to employ efficient algorithms able to keep up with the data flow and to identify events above a given signal-to-noise ratio threshold. The candidate events (or ``triggers'') are then followed up with a range of techniques to decide whether or not a GW signal is present.

Most of the effort on the data analysis side has so far been devoted to the development and implementation of techniques for the mass analysis of the data. These algorithms have been reaching maturity and are being applied to an increasingly large volume of data~\cite{BNS-S2:2006,BBH-S2,HaloBH-S2,S3S4Inspiral,BHNS-S3-spins}; however comparatively less experience has been gained on follow-up methods, see {\em e.g.} Section VIII of Ref~\cite{BNS-S2:2006} (and references therein) and Ref~\cite{followup}. So far the approach to candidate follow-ups is based~\cite{BNS-S2:2006, Allen:2005,Pai-et-al:2001,Chatterji:2006,followup} on (i) addressing the probability of false alarm of an event against the background, usually quantified by repeating the analysis on data from multiple interferometers shifted in time by an unphysical amount(so-called "time-slides"),
(ii) looking for possible correlations between monitoring channels and the GW channel that could reveal an anomalous behaviour of the instrument (or sub-system) or anthropogenic causes around the time of the candidate and (iii) checking that the signal at different interferometers shows consistent behaviour. In this paper we propose a conceptually and practically very different approach to the follow-up of candidates: Bayesian model selection (or hypothesis testing) that is based on the evaluation of the {\em marginal likelihood} or {\em evidence} of a specific model and on the odds ratios between competing hypotheses. Note that in principle the method discussed here could also be used to search for signals in the whole data set; however the computational costs involved in adopting such an approach and the performance of the existing analysis algorithms do not strongly support an effort in this direction at present. 

The key issue that we address  in this paper is, given a data set and a set of prior information, how we calculate the probability of a GW signal being present. In the formalism of Bayesian inference this is translated into considering two models -- Model 1: ``there is a GW signal and noise'', and Model 2: ``only noise is present'' -- and computing the probability of each of them. Bayesian inference provides a conceptually straightforward prescription to evaluate the probabilities, or the odds ratio of the two models. In the context of
GW observations, Bayesian inference is starting to be considered~\cite{ClarkEtAl:2007,ClarkEtAl:2007a,SearleEtAl:2007,UmstaetterTinto:2007,Cannon:2008} as a powerful approach for ground-based observations. The heavy computational load involved in this method when using exhaustive integration has however limited its use on real data. Recently applications of reversible jumps Markov-Chain Monte-Carlo methods (MCMC)~\cite{mcmc} have been considered to tackle this issue in the context of searches for binary inspirals~\cite{UmstaetterTinto:2007}: in this approach Markov-chains are free to move between models, and therefore one can estimate the Bayes factor from the relative time spent by the chains in each one. This technique is promising, still the computational burden is quite significant. In this paper we consider a different and efficient numerical implementation of the direct calculation of odds ratios that makes this approach realistically applicable to extensive follow-up studies of triggers and that we show is robust against a selection of noise artefacts.

The paper is organised in the following way: in Section~\ref{s:modelselection} we introduce the key concepts about
model selection and hypothesis testing in the Bayesian framework; we also introduce the simple case -- possible
observations of inspiralling binary systems in a single interferometer -- that we will consider in the paper;
Section~\ref{s:results} contains the key results of this paper, where we show the power and robustness of 
Bayesian hypothesis testing for follow-up studies; Section~\ref{s:concl} contains our conclusions and pointers to 
future work.

\section{Model selection}
\label{s:modelselection}

\subsection{Overview}
\label{ss:overview}

Let us consider a set of hypotheses (or models) $\{H\}$ and denote with $I$ all relevant prior information we hold.
The predictions made by a model depend on a set of 0 or more parameters, and the possible combinations of parameters define the parameter space of that model. A parameter vector $\vec{\theta}$ represents a point in parameter space, and although each model has its own set of parameters, for ease of notation we shall use $\vec{\theta}$ for them all, and we represent the data under consideration by $d$.

A straightforward application of Bayes theorem yields the probability of a given model:
\be
P(H_j | d, I) = \frac{P(H_j | I)\,P(d | H_j, I)}{P(d | I)}.
\label{e:posteriorMj}
\ee
If the models under consideration are exhaustive and mutually exclusive, then the probabilities above are clearly normalised and satisfy: $\sum_{j = 1}^N P(H_j | d, I) = 1$.
In Equation~\ref{e:posteriorMj}, $P(H_j | d, I)$ is the \emph{posterior probability} of model $H_j$ given the data, $P(H_j | I)$ is the \emph{prior probability} of hypothesis $H_j$ and $P(d | H_j, I)$ is the \emph{marginal likelihood} or \emph{evidence} for $H_j$ that can be written as:
\ba
P(d | H_j, I) & = & {\cal L}(H_j)
\nonumber\\
& = & \int d\vec{\theta}\, p(\vec{\theta} | H_j, I)\, p(d | \vec{\theta}, H_j, I),
\label{e:globalL}
\ea
where $p(\vec{\theta}|H_j,I)$ is the prior probability density distribution for the parameter vector $\vec{\theta}$ and is normalised to 1. $p(d|\vec{\theta},H_j,I)$ is called the \emph{likelihood}, and is an un-normalised measure of the fit of the data to the model.

If we had an exhaustive set of models, we could simply calculate the probability of each model and compare them to find the most likely. Unfortunately we do not have a complete set of models of the data from a GW detector, but we still want to compare the models we \emph{do} have. This is a normal procedure in Bayesian inference, which gives what is termed the \emph{odds ratio} between hypotheses, which is a quantification of their relative probability.

For example, the odds of model $H_j$ against model $H_k$ is
\ba
{\mathrm O}_{jk} & = & \frac{P(H_j|d,I)}{P(H_k|d,I)}=\frac{P(H_j|I)}{P(H_k|I)}\frac{P(d|H_j,I)}{P(d|H_k,I)}
\nonumber \\
& = & \frac{P(H_j|I)}{P(H_k|I)} B_{jk}
\label{e:oddsjk}
\ea
where ${P(H_j|I)}/{P(H_k|I)}$ encodes the ratio of the prior state of belief of the two models and $B_{jk} \equiv {P(d|H_j,I)}/{P(d|H_k,I)}$ is known as 
the \emph{Bayes factor}; the marginal likelihoods $P(d|H_{j,k},I)$ are given by Eq.~(\ref{e:globalL}).  The practical advantage of considering the odds ratio~(\ref{e:oddsjk}) is the fact that
one does not need to evaluate $P(d | I)$.
Of course, if one wanted to evaluate the odds of model $H_j$ 
with respect to all the other independent alternatives, the appropriate quantity is:
\be
{\mathrm O}_{j,\mathrm{other}} = \frac{P(H_j|d,I)}{\sum_{k\ne{}j}P(H_k|d,I)} = \frac{1}{\sum_{j\ne k}O_{kj}}.
\ee
An interesting consequence of using Bayesian model selection
  is that it automatically includes a quantitative version of
  ``Occam's Razor'', the principle that the simpler model should be
  prefered. Because the evidence is found by integrating over the
  entire prior domain of a model, and this prior is normalised to
  unity, the larger the volume of parameter space which the model
  spreads its prior over the lower the resulting evidence will be, all
  else being equal. If two models can assign the same maximum
  likelihood to some data by fitting it equally well, the one which
  makes the more precise prediction of the parameters which produce
  the maximum likelihood will benefit the more than one which makes
         the broader prediction. \cite{UmstaetterLISA,Jaynes}

\subsection{A concrete example}

The specific problem that we are addressing here is how to answer the question: \emph{what are the odds that there is a signal present in the given observations}? We actually need to be more specific and spell out the whole set of background information $I$
for the problem at hand. Here we concentrate on searches for inspiralling compact objects, though the method
outlined here has the potential of much wider applications. The background information $I$ is therefore as follows: (i) the
data set $d$ consist of the superposition of noise $n$ and (possibly) an inspiral GW signal $h$; (ii) if present, only one
GW signal is present at any one time; (iii) the waveform model is exactly known, though the actual parameters 
characterising the source are unknown, within some prior range; 
(iv) gravitational radiation and noise are statistically independent; (v) the noise
is a Gaussian and stationary random process with zero mean and variance (at any frequency) described by a 
\emph{known spectral density}; (vi) observations are carried out with a single instrument.

We are therefore considering a situation in which (schematically) there are only two models:
\begin{itemize}
\item The first hypothesis -- that we will label as $H_\mathrm{S}$, for noise and GW signal -- corresponds to
 ``there is a signal and noise present''; the data set in the time domain is therefore described by
\be
d(t) = n(t) + h(t).
\label{e:d=nh}
\ee
\item The second hypothesis -- that we will indicate as $H_\mathrm{N}$, for noise only -- corresponds to "there is only noise present", and the data set is therefore:
\be
d(t) = n(t).
\label{e:d=n}
\ee
\end{itemize}
The goal of the analysis is therefore to compute the odds ratio, Equation~(\ref{e:oddsjk}), between $H_\mathrm{S}$ and $H_\mathrm{N}$ that we will indicated with $\mathrm{O}_\mathrm{SN}$.

In this paper, for simplicity we consider the observations of gravitational radiation produced during the
inspiral phase of a binary system of non-spinning compact objects. The two objects have individual mass $m_{1,2}$,
and the binary's chirp and total mass are therefore ${\mathcal M}=(m_1{}m_2{})^{3/5}(m_1+m_2)^{-1/5}$ and $m = m_1 + m_2$, respectively. We have also assumed $m_1=m_2$. We model radiation at the leading
Newtonian quadrupole order. As the analysis is more conveniently carried out in the frequency domain,
the model that we adopt is the stationary phase approximation to the radiation in the Fourier domain, see {\em e.g.}~\cite{CutlerFlanagan}. Describing with 
$\tilde g(f)$ the Fourier transform at frequency $f$ of the time-domain function $g(t)$ we can express the
GW signal as:
\be
\tilde h(f;\vec{\theta}) =
\left\{
\begin{array}{ll}
A\,{\mathcal M}^{5/6}\,f^{-7/6}\,e^{i\psi(f;\vec{\theta})}
&  \quad \quad (f \le f_\mathrm{LSO})
\\
0 & \quad \quad (f > f_\mathrm{LSO})
\end{array}\,,
\right.
\label{e:hf}
\ee
where 
\be
\psi(f;\vec{\theta})=2\pi{}ft_c-\phi_c-\frac{\pi}{4}+\frac{3}{4}\left(8\pi{}{\mathcal M}f\right)^{-5/3}
\label{e:psi}
\ee
is the signal phase in the frequency domain and $f_\mathrm{LSO}$ is the frequency of the last stable orbit; in this paper we set $f_\mathrm{LSO}$ equal to the last stable circular orbit for the Schwarzschild space-time and equal mass non-spinning objects, so that
\be
f_{LSO}=(6^{3/2}2^{1/5}\pi{}{\mathcal M})^{-1}\,.
\label{e:flso}
\ee
In Eqs~(\ref{e:hf}) and~(\ref{e:psi}), $\vec{\theta}$ is the 4-dimensional parameter vector describing the GW signal; as parameters we choose
\be
\vec{\theta} = \{A, {\cal M}, t_c, \phi_c\},
\label{e:lambda}
\ee
where $A$ is an overall amplitude that depends on the source position in the sky and
the inclination and polarisation angle of the source, and $t_c$ and $\phi_c$ are the 
time and phase at coalescence. Note that we are using geometrical units in which 
$c = G = 1$ and therefore $M_\odot=4.926\times{}10^{-6}\,$s.

We model the noise as a Gaussian and stationary random process with zero mean, and variance at frequency $f_k$ given by $\sigma_k^2=\left(\frac{f_k}{f_0}\right)^{-4}+1+\left(\frac{f_k}{f_0}\right)^2$,
which roughly represents the shape of a first generation interferometer noise curve up to a multiplicative constant of the order of $10^{-44}$, which is irrelevant as it cancels in the likelihood function;
$f_0 = 150$ Hz is chosen to pick the frequency of maximum sensitivity.

In a real application, the variance of each point can be estimated from the one-sided power spectral density, 
calculated with Welch's method (for example) from the data around the segment of interest~\cite{Welch}.
Notice that in the next section we will consider deviations 
of the actual noise affecting the measurements (but the noise model used to
construct the likelihood will remain unchanged) from the Gaussian and stationary noise
shown above. We will discuss this in more detail in Section \ref{s:results}.

\subsection{Evaluation of the odds ratio}

We can now spell out the odds ratio that needs to be calculated; from Equation~(\ref{e:oddsjk}), this is given by
\be
{\mathrm O}_\mathrm{SN}=\frac{P(H_\mathrm{S}|I)}{P(H_\mathrm{N}|I)}\frac{P(d|H_\mathrm{S},I)}{P(d|H_\mathrm{N},I)}\,.
\label{e:Ogwn}
\end{equation}
For the noise model in this case, there are no free parameters since the noise profile is known, and the evidence is given for Gaussian noise by,
\begin{equation}
P(d|H_\mathrm{N},I)=\prod_{k=1}^N\left(2\pi{}\sigma_{k}^2\right)^{-1}\exp{\left(-\frac{\left|\tilde{d}_k\right|^2}{2\sigma_k^2}\right)},
\end{equation}
where the index $k$ is a short-hand representation for those quantities dependent on frequency $f_k$, and $N$ is the total number of data points. 
Each point $k$ has a real and imaginary observation, the variance of each part will be equal in any realistic data, $\sigma^2_k = \sigma^2_{k{\mathcal R}} = \sigma^2_{k{\mathcal I}}$, although in principle they could differ.
Evaluating the evidence for this case requires no integration.

For the signal model however, the evaluation requires integration over the prior domain (denoted ${\bf \Theta}$) of all parameters, codified in a vector $\vec{\theta}$ and is given by
\begin{eqnarray}
P(d|H_S,I)&=&\int_{\bf \Theta}d\vec{\theta}\,p(\vec{\theta}|H_S,I)\prod_{k=1}^N\left(2\pi{}\sigma_{k}^2\right)^{-1} \nonumber \\
&\times{}&\exp{\left(-\frac{\left|\tilde{h}_k(\vec{\theta})-\tilde{d}_k\right|^2}{2\sigma_k^2}\right)}.\label{eqn:Likelihood}
\end{eqnarray}

The prior probability density function $p(\vec{\theta}|H_S,I)$ is for this proof of concept case uniform on all four parameters, within the domain defined below. It should be noted that the prior must be normalised to one when computing the Bayes factor, and that increasing the range of the prior will decrease the Bayes factor, as the model becomes less predictive and is penalised in automatic accordance with Occam's razor.

In the case of the Newtonian inspiral model, the computation of the marginal likelihood~(\ref{eqn:Likelihood}) requires the evaluation of a 4-dimensional integral which could be calculated using a grid-based approach. However we are interested in developing a general and flexible approach, and for inspiral binaries we will need to expand the evidence calculation to include more realistic waveforms for the analysis of real data with post-Newtonian effects and spins, which have many more parameters: the integral in Eq.~(\ref{eqn:Likelihood}) rapidly becomes unfeasible to integrate exhaustively. To avoid this problem, we have used a probabilistic algorithm called Nested Sampling \cite{Skilling:web,Skilling:AIP,Sivia}, which has been used before in the context of Bayesian hypothesis testing in cosmology~\cite{CosmoNS}, but not considered so far in gravitational wave data analysis. The application of this technique has required significant development and tuning of the algorithm; the details of this work are beyond the scope of this paper and will be documented in a separate publication~\cite{VV08}. Here we shall focus on the advantages of using the evidence in the context of inspiral analyses, which in the next section will be presented with discussion of results from some fiducial data analysis problems.

\section{Results}
\label{s:results}

 In this Section, we will document the results of four particular problems of interest on which the algorithm has been tested: (i) A test on pure Gaussian noise, (ii) the same noise with added signals of different strengths introduced in order to explore the sensitivity of the algorithm to different signal-to-noise ratios. In addition to these cases of obvious relevance, we have also considered the situation where some noise artefacts -- not included into the model for the computation of the marginal likelihood and the Bayes factor -- were added to the data (that is to the simple Gaussian and stationary noise contribution). In particular, (iii) a decaying sinusoid waveform which might approximate an instrumental ringdown, and (iv) a Poissonian noise component. The two latter cases in particular are intended to look at the situation where the models to be tested do not fit the data well at all, in other words they are not exhaustive. This is of interest because in reality there will be more types of interference with the data than we could possibly hope to model, so robustness against such artefacts is an essential characteristic of a search algorithm. This is directly related to the false-alarm rate of existing searches.

In this paper, we have used synthetic data sets in the frequency band 40\,-\,500\,Hz, and used 30\,s of data. The priors defining the range of the model were such that $A\in[0,2\times10^6]$, ${\cal M}\in[7.7,8.3]\,\Ms$, $t_c\in[19.9,20.1]\,s$, $\phi_c\in[0,2\pi)$. The large amplitude reflects the conversion of mass into seconds in Equation~(\ref{e:hf}).  The injected value ${\cal M}=8.0\,\Ms$ was chosen purely to speed up the analysis, as its highest frequency $f_{LSO}({\cal M}=8.0\,\Ms)=478.45\,$Hz allowed frequency components above this value to be eliminated from the innermost loop of the calculation. These prior ranges were chosen to approximate the size of the parameter space that would have to be searched in a real application, which we suggest would be run on candidates triggered by a higher stage in a pipeline, providing some information about the possible characteristics of the signal, namely ${\mathcal M}$ and $t_c$. On these data sets and with such prior ranges, the calculation of the evidence took between several minutes and several hours on a single 2 GHz CPU; the efficiency of the algorithm varies however with the specific tuning in a complicated fashion which will be described in a future paper~\cite{VV08}.  The speed of the computation makes this approach amenable to extension to waveforms characterised by a much larger number of parameters, and to full lists of triggers generated by an inspiral search pipelines, see {\em e.g.} Refs.~\cite{BNS-S2:2006,BBH-S2,HaloBH-S2,S3S4Inspiral}.

\subsection{Sources of Uncertainties}
\label{ss:uncertainties}

Before presenting the results of the example trials, we shall emphasise that there are two contributions to the distribution of results which are obtained when using this method on any data set. These are the inherent uncertainties from the probabilistic nature of the algorithm itself, and the variations in results due to different noise realisations, produced by the random nature of the noise.

The contribution from the noise is an inherent part of any search and the results will naturally vary from dataset to dataset depending on the particular observations made. This cannot be averted, nor should it be. The uncertainty inherent in the algorithm itself however, is a true ``error'' in the result, since there is some particular number which represents the odds ratio, and any deviation from this is undesired. Fortunately, by increasing the running time of the algorithm, these errors can be minimised to a level similar to, or below the variation due to different noise realisations. The inherent uncertainty can in principle be made arbitrarily small, but in the interest of keeping a low run-time, in this paper we have chosen a level roughly similar to the changes caused by the random fluctuations of the noise; this is appropriate for the issues discussed here, and none of our conclusions are affected by the approximation in the evaluation of the integral in Eq.~(\ref{eqn:Likelihood}).

\begin{figure}
\begin{center}
\resizebox{\columnwidth}{!}{\includegraphics{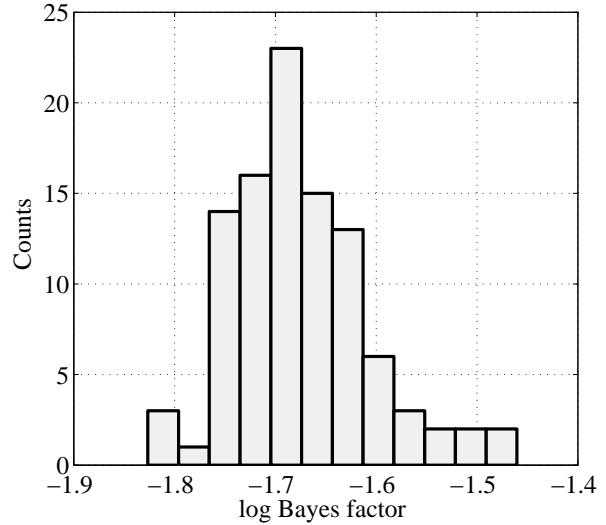}}
\caption{\label{fig:0noises} The distribution of $\log_{10}$ Bayes factors recovered from running the algorithm on 100 different Gaussian and stationary noise realisations, each of length 13\,800 complex samples as described in the text. The distribution in values arises from random fluctuations of the noise, which is contrasted with Figure \ref{fig:0samenoise}, where the inherent uncertainty in the algorithm is shown.
In both cases, the variation in Bayes factors found in datasets with no signal is extremely small compared to the change in Bayes factor that the presence of a signal produces, which is shown in Figure \ref{fig:gaussiansignal} for a range of signal-to-noise ratios.}
\end{center}
\end{figure}

\begin{figure}
\begin{center}
\resizebox{\columnwidth}{!}{\includegraphics{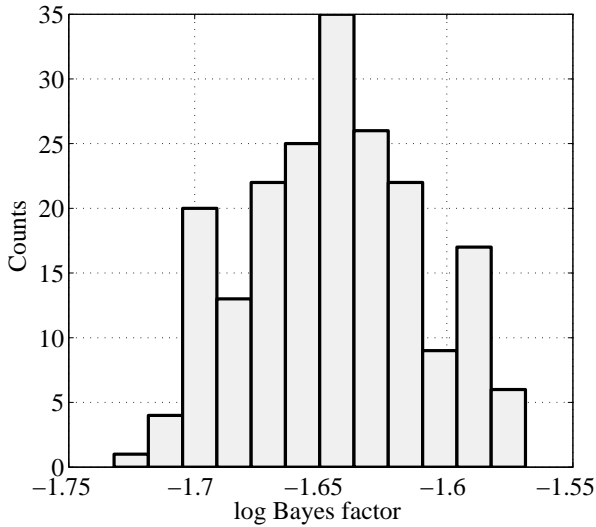}}
\caption{\label{fig:0samenoise} The distribution of $\log_{10}$ Bayes factors recovered from running the algorithm 200 times on the same dataset which contained no signal. The noise is modelled as a Gaussian and stationary random process. The distribution of results is caused by the probabilistic nature of the algorithm, and the range of this distribution can be reduced to have an arbitrarily small range at the expense of increased computation time. In this example and throughout the rest of the paper, the distribution width is chosen to be the same order as the uncertainty caused by different noise realisations, shown in Figure \ref{fig:0noises}.}
\end{center}
\end{figure}

We have run two simple test cases to illustrate the different sources of uncertainties that affect the results; in both cases we have run the algorithm a number of times on signal-free data sets; in the first instance, data sets were generated to contain {\em different realisations} of Gaussian and stationary noise; in the second, the evidence evaluation was performed multiple times on the {\em same} noise realisation; Figures~\ref{fig:0noises} and~\ref{fig:0samenoise} summarise the results.
Figure \ref{fig:0noises} shows the distribution of $\log_{10}$ Bayes factors recovered from the analysis of 100 different noise realisations, and represents the background level of changes in the odds which is caused by random fluctuations in the noise. Figure \ref{fig:0samenoise} shows the inherent uncertainty caused by the algorithm when repeatedly running on a single noise realisation of the same characteristics. The distribution is shown for 200 runs, and the accuracy level has been chosen such that the errors from this effect are not greater then those in Figure \ref{fig:0noises}.

\subsection{\label{ss:nosig}Stationary Gaussian noise with no signal}

The most basic test of whether the evidence-based approach can be helpful in a follow-up analysis is to find how it responds in the absence of a signal. At the present time, this is still the routine situation in gravitational wave data analysis. The results shown in Figure \ref{fig:0samenoise} are the values of the Bayes factor, cf. Eqs~(\ref{e:oddsjk}) and~(\ref{e:Ogwn}), between the signal hypothesis and the noise hypothesis, as calculated by performing 200 runs on identical noise realisations.

The recovered Bayes factors for the hypotheses favour, as expected, the noise only hypothesis, but are (relatively) close to unity. This number is then multiplied by a prior odds ratio, see Eq.~(\ref{e:Ogwn}), to get a total posterior odds ratio.
A reasonable prior would give much larger credence to the hypothesis that no signal is present (i.e. have a value $\ll 1$), in order to reflect the fact that at present sensitivity, we expect inspiral binaries to be rare events. The magnitude of this prior ratio is a factor which quantifies the scepticism of the analysis toward the signal hypothesis, in effect creating a threshold Bayes factor above which the observed signal dominates the prior disbelief.
This number can also be chosen from a procedural point of view, so as to obtain a desired ``false alarm rate'', by performing simulations where known GW signals are added to the data either in hardware or in software.

In the next Section we test the response of the algorithm to a GW inspiral signal added to noise, using signal-to-noise ratios ranging from 1 to 7. From this we shall see that the evidence is an extremely sensitive indicator of the presence of a signal, indeed it is provably the optimal inference for model comparison \cite{Jaynes}.

\subsection{Signal injected into stationary Gaussian noise}\label{Gaussian}

\begin{figure}
\resizebox{\columnwidth}{!}{\includegraphics{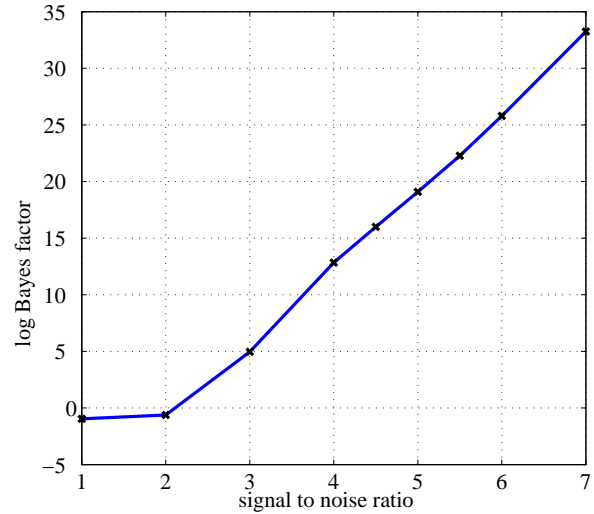}}
\caption{\label{fig:gaussiansignal}The mean recovered $\log_{10}$ Bayes factors from adding an inspiral signal of increasing signal-to-noise ratio, shown on the $x$-axis, to Gaussian and stationary noise. The algorithm was run on each dataset (identical noise realisation) 20 times: each point on the plot corresponds to the mean of the Bayes factor. Note that the Bayes factor grows exponentially with increasing SNR, such that it is an extremely sensitive indication of the presence of a signal. The signal and data parameters were chosen as described in the text.}
\end{figure}

In this section we discuss the results obtained by adding GW inspiral signals, with signal-to-noise ratio (SNR) varying between 1 and 7, to stationary and Gaussian noise, of the same statistical properties as in section \ref{ss:nosig}. The three non-amplitude parameters were kept the same for all of the injections, and had the values ${\mathcal M}=8.0\,{\mathrm M}_\odot$, $t_c=20.0\,$s, $\phi_c=0.0$. The algorithm was run on each of these datasets 20 times, and the results are summarised in Figure \ref{fig:gaussiansignal}. The optimal signal-to-noise ratio~\cite{CutlerFlanagan} is given by  

\begin{equation}
{\mathrm {SNR}}=\sqrt{\sum_{k=1}^N\frac{|\tilde{h}_k|^2}{\sigma_k^2}}\,.
\end{equation}
 
It is very clear from Figure \ref{fig:gaussiansignal}, that the 
Bayes factor  $B_{\mathrm SN}$ rapidly increases with SNR, hence the logarithmic scale on the vertical axis. This shows that the odds ratio between the two candidate models in this case climbs very rapidly to favour the signal model $H_S$ once sufficient evidence is present.

We can now see the effect that the prior odds ratio has on the 
 posterior odds ratio: it acts as a threshold value, above which the overall odds are in favour of the signal model. The value of the prior odds effectively sets a limit on SNR above which the odds favour the signal model, but once this threshold has been reached, the increasing Bayes factor will rapidly climb by many orders of magnitude.

In the cases discussed above, the SNR is chosen by changing the overall amplitude of the injection, while holding the injected chirp mass constant between each case. It should be noted that there is an additional change in the odds ratio which comes about when the injected mass changes. This is due to the fact that the noise evidence $P(d|H_{\mathrm N},I)$ is dependent on the specific shape of the waveform that is  present in the data 
and its inner product with the individual noise realisation. Writing the evidence for the noise model when the data is composed of noise and signal, $\tilde d_k = \tilde n_k + \tilde h_k$, shows the origin of the effect:
\begin{eqnarray}
P(d|H_{\mathrm N},I)&\propto&\prod_k\exp\left[-\frac{1}{2}\left(\frac{\left|\tilde n_k\right|^2}{\sigma_k^2}\right.\right. \nonumber \\ 
&+&\left.\left.\frac{|\tilde h_k|^2}{\sigma_k^2}+\frac{\tilde n_k^* \tilde h_k +\tilde n_k \tilde h_k^*}{\sigma_k^2}\right)\right]
\end{eqnarray}
%
The first term in bracket is a measure of the fit of the noise to the estimated background noise spectrum $\{\sigma_k\}$ independent of the signal. The second term is the signal-to-noise ratio squared, which is independent of the noise realisation. The third term is the important one in this effect: it measures the contrast between the signal waveform and the noise realisation. It is possible to have a constant SNR while the Bayes factor changes if this term varies, although  because the noise and signal are uncorrelated ($\langle \tilde n_k^* \tilde h_k\rangle= \langle \tilde n_k \tilde h_k^*\rangle = 0$), it should be small in comparison to the other terms in this equation. This term can change through either differing noise realisations, or a change in the waveform's \emph{shape}, the phase, or the overall amplitude of the signal. Since we have used the same value of ${\mathcal M}$ in these simulations, this effect should be further reduced.

The choice of the prior odds is an interesting question which we do not attempt to answer conclusively here, but at least two possibilities seem reasonable. Since the prior hypothesis probability $P(H_S|I)$ on the signal model reflects our knowledge before we examine the dataset, we could consider our knowledge of the rate of inspiral events which would be visible to the interferometer in question (see {\em e.g.}~\cite{rates} and references therein), and the length of the prior window on $t_c$. Multiplying these would give us an estimate of the number of inspirals we would expect to see in that time period and mass range, which could be used as the prior odds on seeing an inspiral event in that time.

Alternatively, one could perform a large number of trials on different data realisations and find the distribution of Bayes factors for ``false alarm'' signals with their frequency. A desired false alarm rate could then be set by choosing the prior odds ratio to be the inverse of the Bayes factor for that false alarm rate,  see {\em e.g.}~\cite{VV-cqg:08}. In real data where it is not known \emph{a-priori} if a signal is present, were the search algorithm extended to coherent multiple interferometer models, this could easily be accomplished by offsetting the data from two or more interferometers, so that any real signal would not appear with phase coherence. This is similar to the approach taken in the existing inspiral analysis pipeline to perform background rate estimation, see {\em e.g.} Refs.~\cite{BNS-S2:2006,BBH-S2,HaloBH-S2,S3S4Inspiral,BHNS-S3-spins}. This choice also has appeal as it would automatically include the effect of noise artefacts present in real data which might cause the odds ratio to increase even in the case where there is no astrophysical signal, partially matching the inspiral model.

A more Bayesian treatment of spurious artefacts might attempt to model them too, and then compare their model evidences to those for the GW signal and noise models to test each hypothesis against the others. However, it is unlikely to be possible to model every single noise artefact that appears in a real interferometer, so it is desirable that the basic algorithm functions well even in the presence of such disturbances. In the following sections we report results of tests of the performance of the algorithm in the presence of some typical (though simulated here) deviations of the noise from the simple Gaussian and stationary behaviour. As we will see, the algorithm is less sensitive to these noise artefacts than the true signal, though the Bayes factor decreases as it should. This result indeed supports the usefulness of evidence-based algorithms as part of the tools for follow-up analyses in searches for GWs in the data of current interferometers.

\subsection{Stationary Gaussian noise with ringdown injected}

One type of artefact which is often encountered in real data is the instrumental ringdown, where some component of the detector is excited and produces damped oscillations in the gravitational wave channel, gradually decaying. We have modelled the resulting signal for the purposes of adding it to the synthetic data set as a generic decaying sinusoid waveform in the frequency domain,
\begin{equation}
\tilde{R}(f;A_R,f_R,t_0,\tau)=\frac{A_Re^{-2\pi{}i(f_R-f)t_0}}{\tau^{-1}+2\pi{}i(f-f_R)}.\label{eqn:SNR}
\end{equation}
The parameters of this signal are amplitude $A_R$, frequency $f_R$, starting time $t_0$, and decay time $\tau$ after which the envelope amplitude falls to $e^{-1}$ of its original value. Using this waveform is equivalent to setting the initial phase of the wave to be 0, but this does not affect the generality of the results. In the simulations presented here, a central frequency $f_R=150\,$Hz was chosen, so that the peak of the ringdown would lie in the sensitive region of the frequency domain range.
Ringdowns with $\tau=1\,$s and 10\,s were both added to the dataset, and the algorithm was run as before to search for the presence of a Newtonian inspiral. We stress again that the instrumental ringdown was not part of the models considered for the computation of the Bayes factor.

\begin{figure}
\resizebox{\columnwidth}{!}{\includegraphics{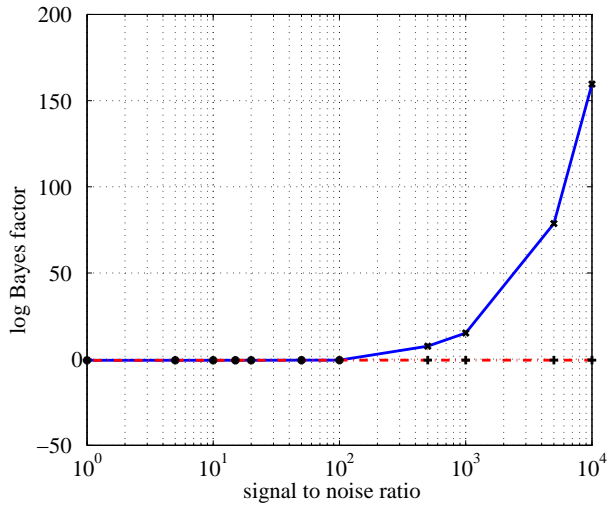}}
\caption{\label{fig:ringdown}The Bayes factors $B_{SN}$ for inspiral model vs. noise model in the presence of an injected ringdown to simulate such events in instrumental noise. Shown here are the results from two ringdowns, with decay times $\tau=1\,$s (dashed line) and $10\,$s (solid line), and varying signal to noise ratios plotted on the $x$-axis. From these results it is clear that the algorithm is robust against interference from this source, as only the $\tau=10$\,s caused an increase in the Bayes factor, and then only at very high SNR.}
\end{figure}

The results of this test are reported in Figure \ref{fig:ringdown}; the signal-to-noise ratio of the instrumental ringdown shown on the $x$-axis is calculated in the same way as for the GW inspiral signal, using Equation \ref{eqn:SNR}, but replacing $\tilde{h}$ with $\tilde{R}$. The figure shows the response of the Bayes factor $B_\mathrm{SN}$ of the inspiral model vs. noise model (for both of which the noise component is modelled as exactly Gaussian and stationary) to a dataset containing Gaussian noise {\em and} an instrumental ringdown signal, which is much more subdued than that to an injected inspiral signal. The vertical axis of the plot of Figure \ref{fig:ringdown} is therefore the same quantity represented on the corresponding axis of Figure~\ref{fig:gaussiansignal}. A ringdown SNR of 1000 is necessary to produce a Bayes factor comparable to that for an inspiral SNR of 4.5 (c.f. figure \ref{fig:gaussiansignal}), when a ringdown with decay constant $\tau=$10\,s is used (for $\tau=$1\,s the effect is even smaller, and hardly noticeable on the scale of the plot). 

\subsection{Stationary and Gaussian noise with additional Poissonian component}\label{s:pois}

\begin{figure}
\resizebox{\columnwidth}{!}{\includegraphics{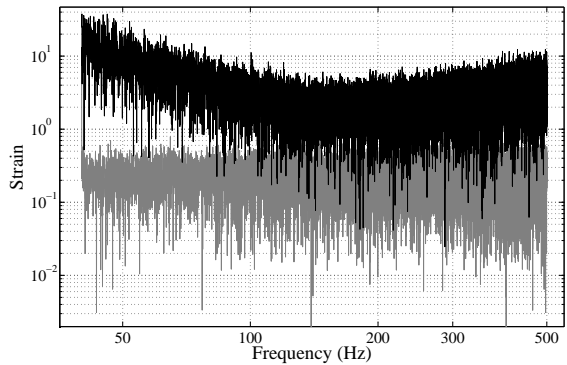}}
\caption{\label{fig:PoisGauss} The amplitude spectra of the Gaussian (upper black) and Poisson (lower grey) contributions to the data used in section \ref{s:pois}.}
\end{figure}

We now explore a different case in which the model's assumption that the noise follows a Gaussian probability distribution is not fully accurate, by injecting an  additional Poissonian noise component into the simulated data.
To accomplish this, a time-series stretch of noise with amplitude drawn from a Poisson distribution in the time domain, scaled by a factor 100 to increase its amplitude, and uniform random phase in the interval $[0,2\pi)$ is generated and Fourier transformed, before being added to the standard Gaussian stationary noise generated in the frequency domain. The root-mean-square of the Gaussian and Poisson components are 4.6 and 0.3 respectively. The spectrum of the Poisson noise showed an approximately uniform power across frequency in contrast with the shaped power spectrum of the Gaussian noise. The amplitude spectrum of the two noise contributions are shows in figure~\ref{fig:PoisGauss}.

The intention of this procedure is to simulate the effects of outliers from the Gaussian noise distribution, which we know occur in the real interferometer data, and which we will be very unlikely to be able to model, making them in effect random. 
In this test we have chosen a Poisson distribution $P(\lambda)$ with a mean $\lambda$ of one point for every ten time stamps, i.e. $\lambda=0.1$. We have explored both the signal-free case and the situation in which a GW signal is added to the (Gaussian + Poissonian) noise, analogous of the studies presented in Secs.~\ref{ss:nosig} and~\ref{Gaussian}, respectively.

\begin{figure}
\resizebox{\columnwidth}{!}{\includegraphics{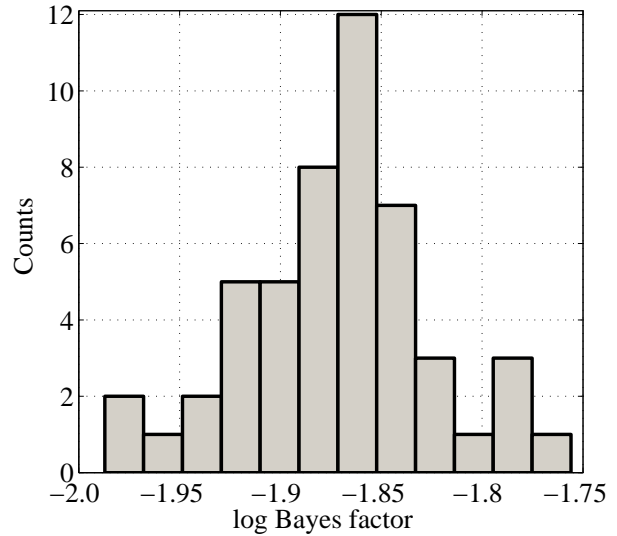}}
\caption{\label{fig:poisnoise} The results from injecting a Poissonian component with distribution $P(0.1)$ into a segment of Gaussian noise (and no GW signal) and running the algorithm 100 times. The presence of Poisson noise does not increase the Bayes factor for an inspiral signal, as the Poissonian noise does not match the template for this waveform. The results of plot should be compared with those obtained on pure Gaussian and stationary noise reported in Figure~\ref{fig:0samenoise}.}
\end{figure}

Figure \ref{fig:poisnoise} shows the results of analysing simply the combination of Gaussian and Poisson noise, without a GW signal present. We can see that the change in Bayes factor due to this additional noise component is minimal (cf. Figure~\ref{fig:0samenoise}), and does not trigger the signal model at all, conversely it depresses the odds. The estimated probability of a signal being present remains low.

\begin{figure}
\resizebox{\columnwidth}{!}{\includegraphics{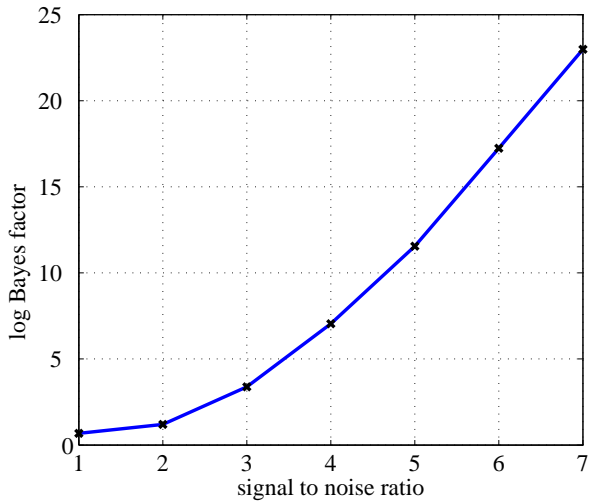}}
\caption{\label{fig:poissig} The results from adding GW signals of varying signal-to-noise ratio onto the combined noise data from a Poissonian and Gaussian distribution, as described in the text. The resulting Bayes factors are lowered in comparison to Figure \ref{fig:gaussiansignal}.}
\end{figure}

Figure \ref{fig:poissig} shows the results of tests similar to those in Section \ref{Gaussian} (GW signal and noise).
In comparison to Figure \ref{fig:gaussiansignal}, the recovered Bayes factors are reduced for the same signal-to-noise ratio -- as an example, $\log_{10} B_\mathrm{SN}\approx  10$ is reached for SNR $\approx 4.6$ instead of $\approx 3.6$ as in the pure Gaussian and stationary noise case -- although the value of SNR for the Poisson dataset does not include the Poisson noise component in its noise estimate, as it is supposed to be a deviation from the noise model.

The results from this test indicate that the presence of an unmodelled yet stochastic noise source will hinder the detection of an inspiral signal, but that the signal model would still be selected, although at a somewhat higher SNR. Such noise will decrease the probability of detection, but it will also decrease the probability of a false alarm with this algorithm as the Bayes factor is always decreased. The exact quantitative consequences for the change of SNR level at which the GW signal model becomes preferred over the noise-only model will depend on the actual character of the deviation from the Gaussian and stationary stochastic process. This can however be rigourously quantified by performing Monte-Carlo simulations on real data. 

Since we do not expect the algorithm to perform \emph{better} when assumptions of the models are violated, it is desirable that it would degrade in the least harmful manner. The results of the tests presented here are therefore reassuring: the Bayes factor will not spuriously generate false alarms in the presence of this type of noise, but instead will simply not perform as well at detecting actual signals. It is worth mentioning again that ideally one would also compute the evidence for this noise model, and that if this was done some of the sensitivity would be restored.

\section{Conclusions}
\label{s:concl}

Bayesian inference using evidence and odds ratios evaluations provides a clear and justifiable means of determining the probability of a signal being present in the data: it is the optimal inference and, as we have shown (although only in simple cases), is also robust against interference from non-gravitational wave signals present in the data, and against deviations in the noise profile. Through the method used in this paper the calculation of odds ratios between hypotheses is made feasible within useful time-scales. The run-time of the code that we have developed can vary depending on factors such as the desired accuracy, whether or not a signal is present, the actual GW waveform and relevant number of parameters that describe the model; however, to perform a single run of one of the analyses typical in this paper takes (much) less than one day on a single 2\,GHz CPU. This is significantly more efficient than other approaches considered so far for Bayesian model selection, {\em e.g.}~\cite{UmstaetterTinto:2007, Littenberg}.  This speed makes it possible to perform thousands of odds ratios calculations per day on Beowulf-type clusters: 
this gives the method a very good combination of sensitivity and speed which we hope will allow the method to be used as one of the techniques for follow-up studies in the analysis of real data.

It is our intention now to further develop this method towards the evaluation of odds ratios in real interferometer data, and integrate it with the existing analyses to provide a robust Bayesian follow-up capable of determining the odds of a signal being present. In order to achieve this, we need primarily to include additional models of GW signals (such as post-Newtonian waveforms) and possibly of instrumental artefacts, which will allow the analysis to distinguish between different types of sources and to eliminate or detect contamination of the noise from unwanted sources.

Another important feature of the method introduced in this paper which has not been discussed here but may be useful in a combined Bayesian analysis is its ability to find the maximum likelihood values of the parameters, which would integrate well with a Markov-Chain Monte-Carlo approach~\cite{mcmc} to full parameter estimation, 
see {\em e.g.}~\cite{RoeverMayerChristensen:2006,RoeverMayerChristensen:2007,Roever-et-al:2007,Slyus-et-al:2007}
and references therein. We intend to include such an interface with MCMC estimation to provide a combined Bayesian follow-up package for inspiral analysis.

\acknowledgments

This work has benefited from many discussions with members of the LIGO Scientific Collaboration. We would like in particular to thank Nelson Christensen and James Clark for fruitful and stimulating discussions. Many of the simulations presented in this paper were performed on the Blue BEAR and Tsunami Beowulf clusters of the University of Birmingham. This work has been supported by the UK Science and Technology Facilities Council. While at Northwestern University, AV was supported by the David and Lucile Packard Foundation and by NASA grant NNG06GH87G.

\end{document}